\documentclass{aa}                   
\usepackage{graphicx}                
\usepackage{txfonts}                 

\newcommand{\kms}{km\,s$^{-1}$}

\newcommand{\kep}{Kepler$-$444}

\newcommand{\teff}{$T_{\rm eff}$}
\newcommand{\tc}{$T_{\mathrm{C}}$}
\newcommand{\ME}{$M_{\oplus}$}

\begin{document}

\title{PEPSI deep spectra.\thanks{Based on data acquired with PEPSI using the Large Binocular Telescope (LBT).
The LBT is an international collaboration among institutions in the
United States, Italy, and Germany. LBT Corporation partners are the
University of Arizona on behalf of the Arizona university system; Istituto
Nazionale di Astrofisica, Italy; LBT Beteiligungsgesellschaft, Germany,
representing the Max-Planck Society, the Leibniz-Institute for Astrophysics
Potsdam (AIP), and Heidelberg University; the Ohio State University; and
the Research Corporation, on behalf of the University of Notre Dame,
University of Minnesota and University of Virginia.}}

\subtitle{III. A chemical analysis of the ancient planet-host star \kep}

\author{C.~E.~Mack~III\inst{1}, K. G. Strassmeier\inst{1}, I. Ilyin\inst{1}, S. C. Schuler\inst{2}, F. Spada\inst{1}, \and S. A. Barnes\inst{1}}

\institute{Leibniz-Institute for Astrophysics Potsdam (AIP), An der Sternwarte
    16, D-14482 Potsdam, Germany; \\ \email{cmack@aip.de}, \email{kstrassmeier@aip.de},
    \email{ilyin@aip.de}, \email{fspada@aip.de}, \email{sbarnes@aip.de} \and University of Tampa, Tampa, FL 33606, USA; \email{sschuler@ut.edu}}

\date{Received ... ; accepted ...}

\authorrunning{C.~E.~Mack~III, K. G. Strassmeier et al.}

\titlerunning{PEPSI deep spectra. III. \kep}

\abstract
{We obtained an LBT/PEPSI spectrum with very high resolution and high signal-to-noise ratio (S/N) of the K0V host  \kep, which is known to host 5 sub-Earth size rocky planets. The spectrum has a resolution of $R$$\approx$250,000, a continuous wavelength coverage from 4230\,\AA\ to 9120\,\AA, and S/N between 150--550:1 (blue to red).}
{We performed a detailed chemical analysis to determine the photospheric abundances of 18 chemical elements, in order to use the abundances to place constraints on the bulk composition of the five rocky planets.}
{Our spectral analysis employs the equivalent width method for most of our spectral lines, but we used spectral
synthesis to fit a small number of lines that require special care. In both cases, we derived our abundances using the MOOG spectral analysis package and Kurucz model atmospheres.}
{We find no correlation between elemental abundance and condensation temperature among the refractory elements (\tc\ $>$ 950 K). In addition, using our spectroscopic stellar parameters and isochrone fitting, we find an age of 10$\pm$1.5 Gyr, which is consistent with the asteroseismic age of 11$\pm$1 Gyr. Finally, from the photospheric abundances of Mg, Si, and Fe, we estimate that the typical Fe-core mass fraction for the rocky planets in the \kep\
system is approximately $24\%$.}
{If our estimate of the Fe-core mass fraction is confirmed by more detailed modeling of the disk chemistry and simulations of planet formation and evolution in the \kep\ system, then this would suggest that rocky planets in more metal-poor and $\alpha$-enhanced systems may tend to be less dense than their counterparts of comparable size in more metal-rich systems.}

\keywords{Stars: abundances -- stars: atmospheres -- stars: late-type -- stars: activity -- planets}

\maketitle

\section{Introduction}

The chemical composition of a planet-hosting star is known to influence the architecture of its planetary system (Dawson \& Murray-Clay 2013; Adibekyan et al. 2013; Beaug\'e \& Nesvorn\'y 2013). Statistical surveys of exoplanet host stars show that Jupiter-like giants are preferentially formed around metal-rich stars (Fischer \& Valenti 2005; Udry \& Santos 2007; Ghezzi et al. 2010) while smaller planets appear to be less selective (Sousa et al. 2011; Buchhave et al. 2012, 2014; Schuler et al. 2015). This has been interpreted as evidence for the prevalence of the core accretion model of planet formation (Ida \& Lin 2004; Mordasini et al. 2009). However, it should be noted that recent work (Nayakshin 2017) has shown that formation by gravitational instability via the tidal downsizing model can also reproduce the observed correlation between giant planets and host star metallicity.

On the other hand, the formation of terrestrial planets does not seem to prefer iron-rich environments and succeeds with similar frequency around stars of various metallicities. Nevertheless, recent studies have hinted at some chemical preference for terrestrial planets. In particular, Adibekyan et al. (2015) showed that metal-poor planet hosts tend to have a higher [Mg/Si] ratio when compared to metal-poor stars without detected planets.
In addition, Dressing et al. (2015) found that, for a small sample of Kepler planets with well known
masses and radii, those planets with masses $<$6\ME\ are likely to have very similar iron to
magnesium silicate ratios, and therefore follow a single mass-radius relationship.
Furthermore, Santos et al. (2015) used a simplified chemical network based on detailed simulations
performed by other teams (Lodders 2003; Seager et al. 2007) to estimate the Fe-core mass fraction
of a subset of the rocky planets from the Dressing et al. sample: Kepler$-$10b, Kepler$-$93b,
and CoRoT$-$7b. Santos et al. (2015) found that their simplified model predicted Fe-core mass fractions
of $\sim$30\% for all three planets. In this paper, we aim to use the same simplified model
to estimate the Fe-core mass fraction for the five rocky planets orbiting the Kepler host star \kep.

The discovery of a multi-planet system around \kep\ (KOI-3158, HIP\,94931) was reported in Campante et al. (2015).
Their photometric analysis revealed five transiting planets with radii between those of Mercury and Venus and orbits within 0.1\,AU of the star (i.e., within the orbit of Mercury). Perhaps even more astounding is the old age of the host star, 11.2$\pm$1.0\,Gyr, which Campante et al. determined from asteroseismology. Furthermore, their spectroscopic analysis of a Keck/HIRES spectrum ($R$$\approx$60,000, S/N$\approx$200) yielded $T_{\rm eff}$=5046\,K and $\log g$=4.6 together with sub-solar abundances of Fe as well as Si and Ti (two $\alpha$-elements) leading to a moderately large [$\alpha$/Fe] index of 0.26\,dex. Thus, it appears that (low-mass) planet formation was already ongoing shortly after the universe was created and that the chemical composition of the pre-stellar material did not have to be metal rich.

\begin{figure*}
{\bf a.}\\
\includegraphics[angle=0,width=\textwidth, clip]{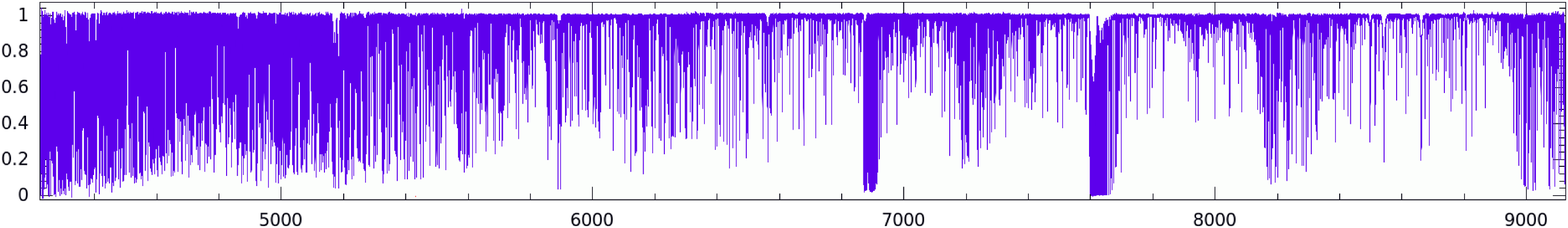}
{\bf b.}\\

\includegraphics[angle=0,width=\textwidth]{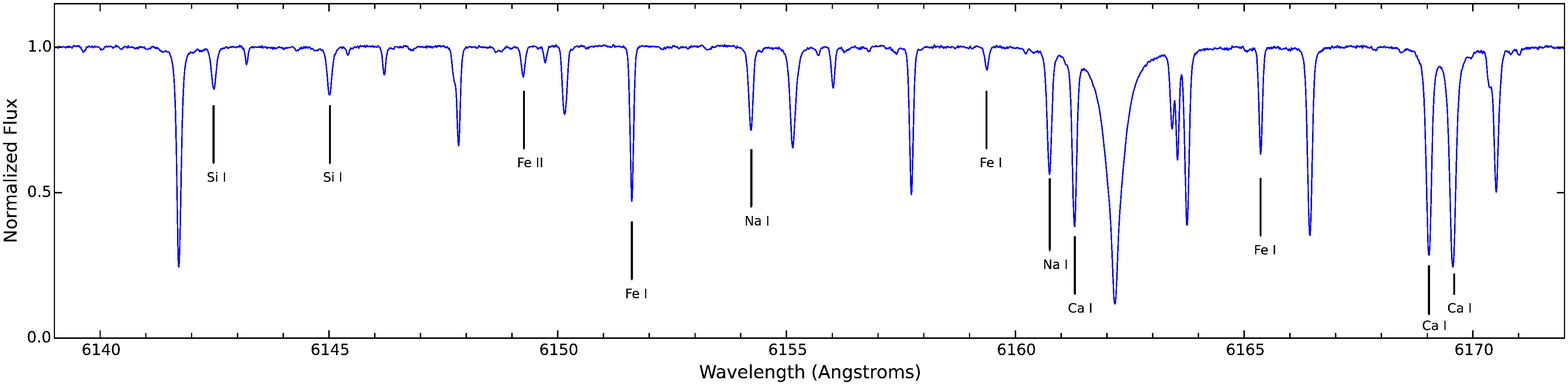}
\caption{\kep\ deep spectrum. \emph{a.} Full spectral region between 4230--9120\,\AA. \emph{b.} Spectral region between $\sim$6140\,\AA\ and $\sim$6170\,\AA. A selection of the lines measured in this region are labeled.}
\label{F1}
\end{figure*}

\begin{table*}
\caption{Spectra observed with PEPSI. } \label{T1}
\begin{tabular}{lllllllllll}
\hline \noalign{\smallskip}
Date        & Time &  Air mass  & Seeing & Exp. & CD & $\lambda$ & S/N$^a$ & CD & $\lambda$ & S/N$^a$ \\
(UT)        &        &               & ('')      & (min) &     & (nm)        &       &      &         (nm)        &              \\
\noalign{\smallskip}\hline \noalign{\smallskip}
2015, May 24        &08:19 & 1.1 & 0.9 & 20 & III &480-544 & 85  &V  &628-742 & 156\\
                    &08:40 & 1.1 & 0.9 & 20 & III &480-544 & 98  &V  &628-742 & 181\\
                    &09:24 & 1.1 & 0.9 & 20 & III &480-544 & 99  &V  &628-742 & 202\\
                    &09:46 & 1.1 & 0.9 & 20 & II  &423-480 & 75  &\dots &\dots & \dots\\
                    &10:14 & 1.1 & 0.9 & 20 & II  &423-480 & 92  &VI &742-912 & 262\\
                    &10:35 & 1.1 & 0.9 & 7  & II  &423-480 & 54  &VI &742-912 & 156 \\
2015, May 25        &03:51 & 1.2 & 0.9 & 20 & II  &423-480 & 98  &IV &742-912 & 177 \\
                    &07:58 & 1.2 & 0.8 & 30 & III &480-544 & 142 &VI &742-912 & 247\\
                    &08:29 & 1.2 & 0.8 & 30 & III &480-544 & 164 &VI &742-912 & 284\\
                    &09:02 & 1.1 & 1.0 & 10 & II  &423-480 & 60  &IV &544-628 & 131\\
                    &10:32 & 1.1 & 1.0 & 30 & II  &423-480 & 133 &IV &544-628 & 261\\
                    &11:03 & 1.1 & 1.0 & 30 & II  &423-480 & 134 &IV &544-628 & 265\\
\noalign{\smallskip}\hline \noalign{\smallskip}
Deep spectrum       & & & & &  I    &\dots     & \dots     & IV &544-628 & 300-500\\
                    & & & & &  II   &423-480 & 150-300 & V  &628-742 & 250-300\\
                    & & & & &  III  &480-544 & 200-300 & VI &742-912 & 550-300\\
\noalign{\smallskip}\hline
\end{tabular}
\tablefoot{$^a$S/N for the individual spectra is given for the central wavelength of each cross disperser (CD). S/N for the deep spectrum is given for the full wavelength range as indicated.}
\end{table*}

In order to perform a detailed chemical abundance analysis of \kep, we obtained high resolution, high signal-to-noise spectra with the optical high-resolution \'echelle spectrograph PEPSI (Strassmeier et al. \cite{pepsi}), which was recently commissioned at the 11.8\,m Large Binocular Telescope (Hill et al. \cite{lbt}). We perform our detailed analysis in order to ascertain the degree to which the photospheric abundances of the host star can place constraints on the bulk compositions of the planets.

In section~\ref{S2}, we describe the observations and the spectrograph, while section~\ref{S3} describes the data analysis. Section~\ref{S4} presents our results, and in section~\ref{S5} we discuss the implications of these results and suggest possible research avenues for future work. Finally, in section~\ref{S6}, we summarize our findings and conclusions.

\section{Observations}\label{S2}

High-resolution $R$$\approx$250,000 spectra of \kep\ were obtained with the Potsdam Echelle Polarimetric and Spectroscopic Instrument (PEPSI; Strassmeier et al. \cite{pepsi}) at the 2$\times$8.4\,m LBT. PEPSI is an asymmetric fiber-fed white-pupil \'echelle spectrograph with two arms (blue and red) typically covering the wavelength range 383--907\,nm. The instrument is stabilized in a pressure and thermally controlled chamber and is fed by three pairs of octagonal fibres per telescope. The different core diameters of the fibres and their respective image-slicer width set the three different resolutions of the spectrograph (43,000, 120,000 and 250,000). The 250,000-mode was used for the spectra in this paper and is made possible with a seven-slice image slicer and a 100-$\mu$m fibre through a projected sky aperture of 0.74\arcsec , comparable to the median seeing of the LBT site. The actual resolution varies with wavelength and depends on the position of the \'echelle order on the CCD and on the focus position. It is generally lowest at the order edges, in particular in the blue arm, and highest near the blaze in the red arm at 7000\,\AA. It ranges from an average of 220,000 in the blue arm to an average of 250,000 in the red arm. The projected image slices are covered in the cross-dispersion direction by 12 pixels per slice. Two 10.3k$\times$10.3k STA1600LN CCDs with 9-$\mu$m pixels record a total of 92 \'echelle orders.

PEPSI can be used either with sky fibers for simultaneous sky and target exposures, or with light from a stabilized Fabry-P\'erot etalon (FPE) for simultaneous fringe and target exposures for precise radial velocities. All spectra in this paper were taken in target+target mode, i.e., no simultaneous sky nor FPE exposures, only two simultaneous target spectra, one from each of the LBT unit telescopes. At the time when the present spectra were obtained, we had just re-positioned the cross dispersers and ended up with an extra \'echelle order in the red-most unit.  Because of (telescope) time constraints we could not use the bluest cross disperser at all. Therefore, our \kep\ spectra cover the wavelength range 4230--9120\,\AA\ with an average dispersion that varies from 7~m\AA/pix in CD\,I to 19~m\AA/pix in CD\,VI.

The present data were taken during spectrograph commissioning and thus full instrument performance was not achieved yet (see paper~I and paper~II; Strassmeier et al. 2018a,b for details). Observations were conducted on May\,23 and 24, 2015. Exposure times of  7, 10, 20, and 30\,min were chosen depending upon wavelength range, seeing, and time availability. Table~\ref{T1} summarizes the observing log.  Basic data reduction followed the procedures laid out and described in detail in paper~I.

After optimal extraction and wavelength calibration the one-dimensional spectra are co-added to build the single ``deep spectrum'' in Table~\ref{T1}. Altogether, the deep spectrum is made from seven individual spectra in CD\,II, five in CD\,III, four in CD\,IV, three in CD\,V, and four in CD\,VI. Each pixel of each spectrum is weighted according to its inverse variance from the optimal extraction and is optimally re-sampled before adding. Also notice that each individual spectrum is actually two spectra in our case because the LBT is in fact a combination of two telescopes. At one occasion on May~24, 09:46UT the DX~M1 mirror cell panicked during integration and the spectrum from one side was lost. Figure~\ref{F1}a shows the full spectrum. A portion of it spanning the wavelength region $\lambda$6140-$\lambda$6170 is shown in Fig.~\ref{F1}b.

\section{Analysis}\label{S3}

The chemical abundances (relative to solar) of 18 elements have been derived from the deep spectrum of \kep. The 2014 version of the LTE spectral analysis package MOOG (\cite{1973ApJ...184..839S}) was used to perform the spectral analysis. The abundances were derived from measurements of the equivalent widths (EWs) of atomic lines using the SPECTRE analysis package (\cite{1987BAAS...19.1129F}). By requiring excitation and ionization balance of the \ion{Fe}{i} and \ion{Fe}{ii} lines, we were able to derive the stellar parameters: \teff, $\log{g}$, [Fe/H], and $\xi$ (microturbulence). The line lists for the wavelength regions encompassing each element were taken from the Vienna Atomic Line Database (VALD; \cite{1995A&AS..112..525P,1999A&AS..138..119K}). Abundances are always given relative to solar\footnote{$[X/H] = \log(N_X)_{\rm star} - \log(N_X)_{\rm Sun}$ with $\log (N_X) = \log (N_X/N_H) + 12$. }. The solar spectrum that we used is the grand average from paper~I (Strassmeier et al. 2018a). 

The carbon abundances were derived from  the C$_2$ Swan system features at $\lambda5086$ and $\lambda5135$ using the \emph{synth} driver in MOOG to synthetically fit them by eye. In addition, the equivalent widths of three high-excitation \ion{C}{i} lines were measured: $\lambda5052.17$, $\lambda5380.34$, and $\lambda7113.18$. Abundances were derived from these equivalent width measurements using the MOOG \emph{abfind} driver.

The two components of the C$_2$ Swan system yield relative abundances that are identical, i.e., [C/H]$=-0.56$. However, the high-excitation \ion{C}{i} lines yield discrepant relative abundances of [C/H]$={-0.31, -0.07, +0.22}$, respectively. Previous LTE analyses of high-excitation lines, such as the \ion{O}{i} triplet at $\lambda$7775\,\AA\ (e.g., Schuler et al. 2004, 2006; Ram\'irez et al. 2013), have found that the abundances derived from high-excitation lines increase with decreasing \teff\ for stars with \teff$\leq$5,400\,K. For the high-excitation \ion{C}{i} lines in particular, Schuler et al. (2015) found that these lines also give abundances that are higher by $\sim$0.16\,dex than the C$_2$ Swan lines for Kepler-37, a star with \teff\ of 5,400\,K. Our C abundances for Kepler-444 seem to follow this same pattern, suggesting that in this cool star, the \ion{C}{i} lines do not provide reliable abundances. We therefore adopt the abundance determined from the C$_2$ Swan system as the carbon abundance here.

\begin{table}
\caption{Stellar parameters and abundances.} \label{tab:params_kep444}
\begin{tabular}{lll}
\hline \noalign{\smallskip}
& \kep & \\
\noalign{\smallskip}\hline \noalign{\smallskip}
	\teff\ (K)             && $5,172 \pm 75$ \\
	$\log g$ (cgs)         && $4.56 \pm 0.18$ \\
	$\xi$ (\kms)           && $1.64 \pm 0.37$ \\
	$[$C/H$]$              && $-0.56 \pm 0.000^{a} \, \pm 0.03^{b}$ \\
	$[$O/H$]^c$              && $-0.11 \pm 0.027 \, \pm 0.12$ \\
	$[$Na/H$]$             && $-0.44 \pm 0.055 \, \pm 0.08$ \\
	$[$Mg/H$]$             && $-0.27 \pm 0.033 \, \pm 0.05$ \\
	$[$Al/H$]$             && $-0.19 \pm 0.014 \, \pm 0.05$ \\
	$[$Si/H$]$             && $-0.37 \pm 0.022 \, \pm 0.04$ \\
	$[$K/H$]$              && $-0.38 \pm \dots  \, \pm 0.13$ \\
	$[$Ca/H$]$             && $-0.53 \pm 0.020 \, \pm 0.10$ \\
	$[$Sc/H$]$             && $-0.34 \pm 0.011 \, \pm 0.08$ \\
	$[$Ti/H$]$             && $-0.27 \pm 0.036 \, \pm 0.11$ \\
	$[$V/H$]$              && $-0.18 \pm 0.024 \, \pm 0.11$ \\
	$[$Cr/H$]$             && $-0.46 \pm 0.006 \, \pm 0.07$ \\
	$[$Fe/H$]$             && $-0.52 \pm 0.011 \, \pm 0.12$ \\
 	$[$Mn/H$]$             && $-0.61 \pm 0.030 \, \pm 0.10$ \\
	$[$Co/H$]$             && $-0.33 \pm 0.015 \, \pm 0.06$ \\
	$[$Ni/H$]$             && $-0.50 \pm 0.013 \, \pm 0.05$ \\
	$[$Zn/H$]$             && $-0.41 \pm 0.004 \, \pm 0.08$ \\
	$[$Y/H$]$              && $-0.51 \pm \dots \, \pm 0.12$ \\
\noalign{\smallskip}\hline
\end{tabular}
\tablefoot{Adopted solar parameters (Cox 2000): \teff\ $=5,777$ K, $\log g=4.44$, and $\xi=1.38$~\kms.
$^a$The uncertainty in the mean. For C, both spectral lines yielded the same abundance; for K and Y, the abundance was derived from a single spectral line; and for Co, all the spectral lines yielded the same abundance, except one that differed by 0.01\,dex. $^b$The quadratic sum of the uncertainty in the mean and the uncertainties due to \teff, $\log{g}$, and $\xi$. $^c$Corrected for NLTE effects, see text. }
\end{table}

For the oxygen abundances, four spectral lines were analyzed: the forbidden line at $\lambda6300.34$ and the near-infrared triplet at $\lambda7771.94$, $\lambda7774.17$, and $\lambda7775.39$. The forbidden line was analyzed by measuring the equivalent width and using the MOOG \emph{blends} driver to account for known blends with spectral lines from other elements, in particular the known blend with an Ni line. For the near-infrared triplet, we measured their equivalent widths and used the standard MOOG \emph{abfind} driver. We then corrected the near-infrared triplet abundances for NLTE effects using the corrections from Takeda (2003). The forbidden line and the near-infrared triplet yielded relative abundances that agree reasonably well: [O/H]=$-0.12$ and $-0.09$, respectively. More recently, Amarsi et al. (2016) provided three-dimensional (3D) NLTE corrections for 1D LTE abundances for the temperature range 5,000-6,500\,K. Applied to our measurements, it corrects to abundances of --0.13 and --0.11 for the forbidden and the near-infrared triplet, respectively. Therefore, we simply average these abundances and adopt an [O/H] abundance of $-0.11$. We note that significant disagreement between the near-infrared triplet and forbidden line relative abundance has been seen before in open cluster dwarf stars with spectral types similar to \kep\ (Schuler et al.\ 2006). However, there is some evidence that the discrepancy between these two [O/H] indicators may decrease with age, which we discuss further in section~\ref{dis_co}.

For V and Co, in order to account for possible hyper-fine structure (hfs) effects (\cite{2000ApJ...537L..57P}),
we used the MOOG \emph{blends} driver on specific spectral lines: the \ion{V}{i} lines at $\lambda$6090.21 and $\lambda$6111.65, and the \ion{Co}{i} lines at $\lambda$5301.04, $\lambda$5352.04, and $\lambda$6814.94. The hfs components for most of these spectral lines were obtained from Johnson et al. (2006), but the components for $\lambda$5352.04 came from Jofr\'e et al. (2017). The solar-normalized abundances for these V and Co lines derived using the \emph{blends} driver are in good agreement with the solar-normalized abundances derived from the V and Co lines that were analyzed with the standard MOOG \emph{abfind} driver. Therefore, the hfs effects for these lines evidently have negligible impact on the solar-normalized abundances, and the adopted V and Co abundances quoted in
Table~2 are the mean of the abundances derived using the \emph{blends} and \emph{abfind} drivers.
We also note that hfs affects mostly lines with larger EWs, which we found to be roughly $\geq$40\,m\AA .

For two elements, K and Y, the abundance was derived from the measurements of a single spectral line. In order to obtain the best estimate of the total uncertainty in the abundance for these particular lines, we estimated how the uncertainty in the EW-measurement affected the uncertainty in the abundance. For simplicity, a minimum-maximum method was used. For both lines, we chose three different continuum placements within the rms scatter of the local continuum, and for each continuum placement made four different measurements of the EW. This gave us a set of 12 EWs for each line in both the \kep\ and solar spectra. For each line, the maximum and mininum EW values were used to compute a maximum and minimum $\log (N_X)$. The differences between the maximum and minimum $\log (N_X)$ were adapted as the uncertainties in $\log (N_X)$ and propagated through the computation of the solar-normalized abundances. These uncertainties in the solar-normalized abundances (due to the uncertainty in the EWs) were then added in quadrature to the errors in the abundances due to the stellar parameters, in order to determine the final uncertainties quoted in Table~2. We also note that the sensitivity of the derived abundances to the adopted metallicity is small compared to the sensitivities to \teff, $\log g$, and $\xi$. 

We also determined the Mn abundance as another odd-$Z$ element. Mn is only weakly NLTE dependent, corrections are only 0.003\,dex (Bergemann \& Gehren 2008), even less than for O and thus we neglected NLTE effects for Mn. However, we have reliable hfs components (Johnson et al. 2006) for only three Mn lines ($\lambda$5432.546, $\lambda$6013.513, $\lambda$6021.819). We measured other Mn lines that we did not have hfs components for and found a large spread of up to --0.2\,dex in the abundances, so we do not consider them to be reliable without hfs corrections. The average of the three lines is given in Table~2. In addition, we attempted to measure Li, but there was no spectral line distinguishable from the continuum noise at $\lambda$6707.8. There were also no discernible spectral lines for the high-excitation N lines at $\lambda$7468.3, $\lambda$8216.3, and $\lambda$8683.4.

The analysis we performed was nearly identical to the analysis of the planet-hosting wide binaries HD\,20782/81 (Mack et al.\ 2014) and HD\,80606/07 (Mack et al.\ 2016). The stellar parameters and relative abundances that we derived
for \kep\ are summarized in Table~\ref{tab:params_kep444}.  The adopted line list, EWs, and line-by-line $\log{N}$ abundances of each element for \kep\ and the Sun are given in the Appendix in Table~\ref{tab:lines_kep444}.

\section{Results}\label{S4}

\begin{figure*}
\centering
\includegraphics[angle=0,width=12cm, clip]{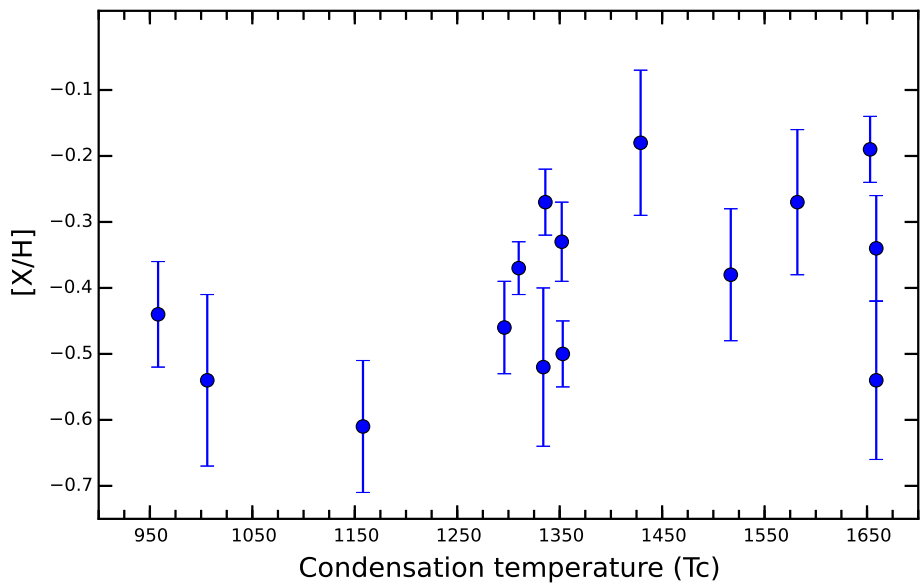}
	\caption{\kep\ relative abundances vs. condensation temperature (\tc $>950$ K).}
	 \label{F2}
\end{figure*}

\subsection{Stellar parameters and abundances}

The stellar parameters (\teff=5,172$\pm$75\,K, $\log{g}$=4.56$\pm$0.18, $\xi$=1.64$\pm$0.37 \kms, and [Fe/H]=$-$0.52$\pm$0.12) and elemental abundances (Table~\ref{tab:params_kep444}) derived for \kep\
are consistent with it being a metal-poor $\sim$K0 dwarf. The $\log{g}$ value is identical to the one derived from
asteroseismology by Campante et al.\ (2015), whereas our value for \teff\ differs from theirs by 126\,K. However, the two \teff\ values still agree within 3-$\sigma$.

We also note here that given the $\approx$50-day rotation period of \kep\ (see section~\ref{gyro}), a radius
of 0.75~$R_{\odot}$ translates into a maximum $v\sin{i}$ of $\approx$0.8 \kms. This value is considerably less than both our value for the microturbulence (1.6~\kms) and the resolution of PEPSI ($\approx$1.3~\kms). Therefore, the spectral lines of \kep\ should not be significantly affected by rotational broadening.

\kep\ has significantly enhanced $\alpha$-abundances. To determine [$\alpha$/Fe] for \kep, we averaged [O/H], [Mg/H], [Si/H], [Ca/H], and [Ti/H] to obtain a value of [$\alpha$/Fe]=0.23. This value agrees well with the value of 0.26 that Campante et al.\ derived using their values for [Si/H] and [Ti/H].

Previous studies (e.g., Mel\'endez et al. 2009, Ram\'irez et al. 2010, Schuler et al. 2011, Mel\'endez et al. 2014) have shown that Sun-like stars tend to show an increase in elemental abundance as a function of elemental condensation temperature, while the Sun itself appears to be depleted in elements with high condensation temperatures, that is the refractory elements. Since it is these refractory elements that are the main constituents of rocky, terrestrial planets, it has been suggested (Mel\'endez et al. 2009, Ram\'irez et al. 2010) that the Sun being depleted in refractory elements may reflect the fact that these elements are locked-up in the Solar System's rocky planets. Therefore, given that \kep\ hosts five rocky planets, we investigated whether its abundances show any particular trend with condensation temperature. 

Figure~\ref{F2} shows the relative abundances of \kep\ versus condensation temperature (\tc) for the refractory elements (\tc $>$950\,K). These relative abundances are the means of the line-by-line differences between $\log{N}_{\rm Kepler-444}$ and $\log{N}_{\rm Sun}$.  For the condensation temperatures, we used the 50\% \tc\ values
tabulated in Lodders (2003). There is no evident trend in the abundances as a function of condensation temperature. A simple linear regression fit returns an insignificant positive slope of +0.00019$\pm$0.00012 dex/K with a standard fit deviation of 0.11 and chi-square of 0.21. A Pearson correlation gives a coefficient of 0.393, which for a sample size of 15, gives an adjusted correlation coefficient of only 0.07. 

In the following subsections, we describe several age indicators that we used to check for consistency with the asteroseismic age of 11$\pm$1 Gyr. We also note that while each of the following age indicators have uncertainties greater than that of the asteroseismic age, each indicator does yield an age ($\sim$10-11\,Gyr) that agrees well with the asteroseismic one, and that our main goal is to show that a variety of age indicators consistently produce an old age for \kep.

\subsection{The Yale-Potsdam Isochrones}

Given the spectroscopic $\log{g}$ and \teff\ we determined for \kep, we used the Yale-Potsdam isochrones (YaPSI;
Spada~et~al. 2017) to determine a mass and age for \kep. We used the YaPSI models in conjunction with the Monte Carlo Markov Chain best-fit search tool BAGEMASS (Maxted et al.\ 2015). The $\chi^2$ minimization was performed with respect to the available observational constraints on the following parameters: $L/L_{\odot}$, \teff, $\log{g}$, and [m/H] (the alpha-enhanced rescaled metallicity; see below). Gravitational settling of He and heavy elements is also taken into account in the YaPSI models; this effect is important for a low-mass, old star such as \kep. Indeed, for \kep, an initial [m/H] of $\approx$--0.30 is required to fit the present value of [m/H]$=$--0.4. Our best-fit solution yields a mass of $M$=0.76\,M$_{\odot}$ and age of 10$\pm$1.5\,Gyr. These values are in good agreement with those derived by Campante et al. (2015).

\begin{figure*}
\centering
\includegraphics[angle=0,width=12cm, clip]{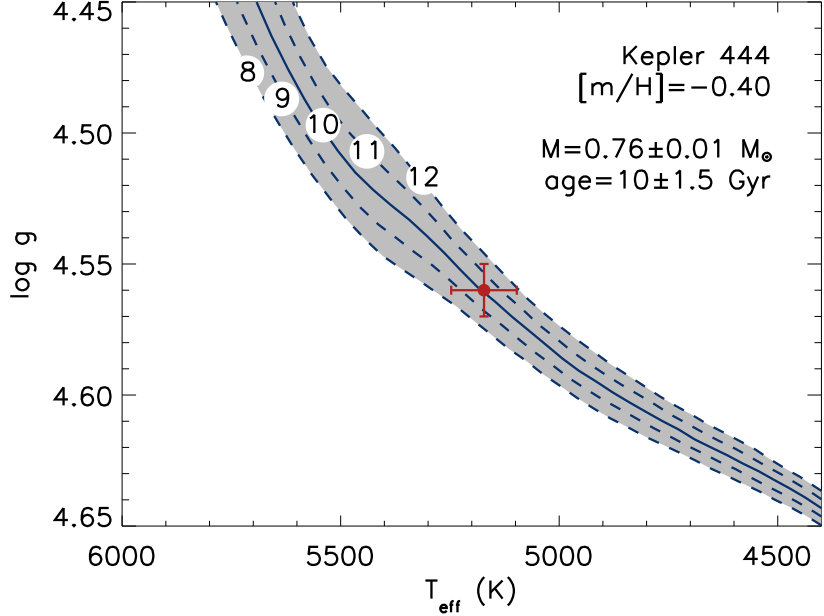}
\caption{\kep\ H-R diagram. The point with error bars corresponds to the \teff and $\log{g}$ of \kep. The curves are the isochrones from YaPSI. The numbers in circles indicate the age in Gyr for each isochrone. The solid curve is the best-fit age.}
	 \label{F3}
\end{figure*}

In order to take into account the enhancement in $\alpha$-elements ([$\alpha$/Fe]$\approx$0.2), we used the scaling relation of Salaris et al. (1993). Essentially, this scaling mimics the effects on the stellar models from the $\alpha$-enhanced abundances by introducing an effective, rescaled metallicity ([m/H]) that is slightly more metal-rich than the spectroscopically determined metallicity ([Fe/H]). These scaling relations are valid for stars that are sufficiently metal-poor (Kim et al. 2002).  In the case of \kep, with [$\alpha$/Fe]$\approx$0.2 and [Fe/H]$=$--0.52, we obtain a rescaled metallicity of [m/H]$=$--0.40; this is the value quoted in Fig.~\ref{F3}. Note that the dashed lines in Fig.~\ref{F3} represent YaPSI isochrones with the same [m/H] for a range of ages that bracket the best-fit value of 10\,Gyr. The shaded gray region is merely intended to guide the eye and highlight the locus defined by these isochrones; it is not a confidence interval.

\subsection{Gyrochronology}\label{gyro}

Mazeh et al. (2015) has determined a rotation period of 49.4 $\pm$ 6 days for \kep\ from Kepler satellite data. Although the uncertainty in the period is large, it is typical for stars of such long periods in Kepler data, given the $\approx$90-day length of each Kepler quarter. The relatively long period immediately indicates a fairly old star.

Stars of comparable mass to \kep\ have rotation periods of $\approx$24 days in the 2.5 Gyr-old open cluster NGC\,6819 (Meibom et al. 2015), and $\approx$31 days in the 4~Gyr-old open cluster M\,67 (Barnes at al. 2016). Extrapolation of the Skumanich-type rotation-age relationship to the period of \kep\ implies an age slightly in excess of 10\,Gyr.

A more precise formulation of gyrochronology, validated by reproducing the mass and age dependencies of rotation in a series of open clusters, including the 4~Gyr-old open cluster M\,67, is available in Barnes (2010). This formulation yields a gyro age for \kep\ of 10.9$\pm$2.5~Gyr, where we have used the convective turnover timescales tabulated in Barnes \& Kim (2010). The uncertainty in the gyro age is completely dominated by the relatively large uncertainty in the rotation period determination.

The gyro age is thus consistent with the asteroseismic determination, but has larger uncertainties in this particular case.

\subsection{The [Y/Mg]-clock}

Also, we measured [Y/H] in order to take advantage of the [Y/Mg] age indicator for solar twins that was introduced by Nissen (2015) and further investigated by Tucci Maia et al. (2015). While \kep\ is certainly not a solar twin, we thought it would be interesting to see if [Y/Mg] would yield an age consistent with the asteroseismic age and the other age indicators we employed. We determined that [Y/Mg]=--0.19 for \kep, which using the relation from Tucci Maia et al. (2015) corresponds to an age of $\approx$9\,Gyr. Given that their [Y/Mg]-age equation is specifically tuned for solar twins, this age estimate is not too discrepant with the $\approx$10-11\,Gyr ages determined from asteroseismology, gyrochronology, and the YaPSI isochrones. Recent studies (Feltzing et al. 2017, Delgado Mena et al. 2017) have shown that thick disk stars with [Fe/H]$<$$-0.5$ do not show a tight correlation between age and [Y/Mg]. Therefore, as a member of the thick disk with an [Fe/H] of $-0.52$, \kep\ may be just beyond the limit of the range of applicability for the [Y/Mg]-clock.

\subsection{The $^{12}\rm{C}$/$^{13}\rm{C}$ isotope ratio}

Finally, we derived an approximate value for the $^{12}$C/$^{13}$C isotope ratio. As a zeroth-order approximation,
we measured the equivalent widths of the CN features at 8003.5\,\AA\ (due to $^{12}$C) and 8004.5\,\AA\ (due to $^{13}$C). It was not possible to fit a simple Gaussian or Voigt profile to these features, so we integrated the area under the line profile using the Simpson's rule integration procedure in MOOG. Repeated adjustments of where the wings of the spectral line merge with the continuum yielded the following equivalent widths for the two lines respectively; 5.2$\pm$0.1~m\AA\ and 3.0$\pm$0.4~m\AA, where the error is simply the standard deviation in the measurements. Taking the ratio of these two values results in a very low isotope ratio of 1.7, which is consistent with \kep\ having an old age.

\section{Discussion}\label{S5}

\subsection{The C and O abundances of \kep}\label{dis_co}

The C and O abundances of \kep\ are consistent with the findings of Bensby \& Feltzing (2005) for stars with similar [Fe/H]. In their Figure~11, it is clear that \kep\ lies near the edge of the distribution of stars with [Fe/H]$\approx$--0.5, but it is certainly within the overall scatter. However, in Delgado-Mena et al. (2010) and Nissen et al. (2014), the C abundance of \kep\ falls below the general distribution of stars with [Fe/H]$\approx$--0.5. Still, given the scatter in the data for each of these studies, and the relatively few stars with metallicities near $-0.5$ (varying from 3--10 data points across the three studies), \kep\ can not definitively be considered an outlier, but it has perhaps a relatively low C abundance compared to the stars with similar metallicities, kinematics, and $\alpha$-enhancement. 

In addition, the lack of discrepancy in the relative oxygen abundances ([O/H]) derived from the [\ion{O}{i}]-6300 forbidden line and the near-infrared triplet is in agreement with the suggestion from Schuler et al. (2006) that this discrepancy disappears with age. In their studies of the oxygen abundances for the Pleiades, M\,34, and Hyades open clusters, as well as for the UMa moving group (Schuler et al. 2004, King \& Schuler 2005, and Schuler et al. 2006), they found unexpectedly large O abundances derived from the near-infrared triplet among the K dwarfs in all three open clusters and in the UMa moving group. However, for the Hyades and the UMa moving group (which both have an age around 600\,Myr), the discrepancy was significantly smaller than in the Pleiades and M\,34 (which have ages in the range 100$-$200\,Myr). Therefore, they posited that the discrepancy may decrease with age, and the nearly identical O abundances derived from the 6300-\AA\ forbidden line and the near-infrared triplet in \kep\ may be the result of this trend. If this trend could be confirmed by studying the O abundances of K dwarfs in a statistical sample of open clusters and moving groups that span a wide range of ages, then this could potentially become a powerful spectroscopic age indicator.

\subsection{How \kep\ compares to other known metal-poor planet hosts}
	
Adibekyan et al. (2012) showed that metal-poor host stars with small planets are enhanced in $\alpha$-elements with respect to metal-poor stars without known planets. Furthermore, Adibekyan et al. (2015) showed that small planets tend to prefer host stars with larger [Mg/Si] ratios. Since \kep\ falls near the upper limits of the distributions shown in Figure~4 of Adibekyan et al. (2015) and Figure~2 of Adibekyan et al. (2012), the enhanced $\alpha$-abundance ([$\alpha$/Fe]=0.23) and the comparatively large value of [Mg/Si]=0.1 for \kep\ are consistent with the findings from Adibekyan et al. (2012, 2015). Thus, \kep\ is a member of the growing population of metal-poor host stars that are enhanced in $\alpha$-elements and possess larger [Mg/Si] when compared to metal-poor non-hosts.

In addition, Adibekyan et al. (2015) suggest that since [Mg/Si] has decreased over time as a result of Galactic chemical evolution, the interior structure and composition of older terrestrial planets, like those in the \kep\ system, might be significantly different from the terrestrial planets in the solar system. Indeed, compared to the solar system, the mass fraction of Fe may have been considerably smaller, resulting in Earth-mass planets with significantly lower densities and correspondingly larger radii. In the next section, we estimate the possible Fe-core mass fraction of the rocky planets in the \kep\ system using the photospheric chemical abundances.

\subsection{Estimating the iron core mass fraction from the photospheric abundances}

Here we follow the procedure and simplified chemical network used in Santos et al. (2015). This network consists of H, He, C, O, Mg, Si, and Fe because equilibrium condensation models (Lodders 2003; Seager et al. 2007) show that these seven elements are the most important for determining the major species of condensates in the proto-planetary disk, which are H$_{2}$O, CH$_{4}$, Fe, MgSiO$_{3}$, and Mg$_{2}$SiO$_{4}$. Using the photospheric abundance ratios as proxies for the disk abundance ratios, we can relate our derived photospheric atomic abundances for C, O, Mg, Si, and Fe to the disk abundances of the five major condensates listed above.  These disk abundances can be converted to mass fractions for each condensate. From these mass fractions the Fe-core mass fraction of a rocky planet can be estimated by dividing the mass fraction of Fe by the combined mass fractions of Fe and the two magnesium silicates. From our calculations, for the rocky planets in the \kep\ system, we would typically expect an Fe-core mass fraction of approximately 24\%.

When interpreting this value for the Fe-core mass fraction, it is important to keep in mind the following caveats:
        \begin{itemize}
           \item[(a.)] This is only a first-order estimate using a simplified chemical network. A full numerical analysis
               using the photospheric abundances as inputs to determine which molecular and atomic species condense out of
               the disk should be performed in order to verify this result.
           \item[(b.)] Given that \kep\ is metal-poor and $\alpha$-enhanced, there may be other important disk condensates that are not
               included in the Santos et al. (2015) simplified chemical network, which is only valid for disk compositions where the Mg abundance is greater than the Si abundance.
           \item[(c.)] From the YaPSI isochrone analysis that we used to determine an age for \kep, we found that due to gravitational settling,
               \kep\ may have initially been $\approx$0.1\,dex more metal-rich. Therefore, the present-day photospheric abundances of \kep\
               may be less accurate proxies of the initial disk composition than is typically assumed for stars with solar-type composition.
        \end{itemize}
Keeping these caveats in mind, this estimate of a typical Fe-core mass fraction of $\approx$24\% for terrestrial planets in the \kep\ system, which is significantly less than the $\sim$30\% Fe-core mass fraction observed in the solar system and predicted for other systems with solar-type composition (Santos et al. 2015), implies that rocky planets around metal-poor, $\alpha$-enhanced host stars may tend to be less dense than rocky planets of comparable size around more metal-rich host stars. Indeed, a recent study (Santos et al. 2017) that estimated the composition of potential planets around a sample of 1,111 stars observed with HARPS (Delgado Mena et al. 2017; Adibekyan et al. 2012), predicts that for old, thick disk stars the Fe-core mass fraction for rocky planets would be 20$-$25\% (priv. comm. V.~Adibekyan).

\section{Summary}\label{S6}

We have performed a detail chemical analysis of the planet-hosting star \kep, using a high-resolution, high S/N spectrum obtained with the recently commissioned LBT/PEPSI spectrograph. We find that \kep\ is $\alpha$-enhanced and metal-poor, in agreement with the initial spectroscopic analysis of Campante et al. (2015). Furthermore, using different age indicators, such as the YaPSI isochrones, gyrochronology, and the [Y/Mg] ratio, we find an age of around 10\,Gyr for \kep, which is consistent with the $11\pm 1$\,Gyr that Campante et al. determined from their asteroseismic analysis.

In addition, we note that the good agreement between the relative (to solar) oxygen abundances determined from the [\ion{O}{i}]-6300 forbidden line and the near-infrared \ion{O}{i}-triplet in \kep\ may point towards a potentially powerful spectroscopic age indicator for K dwarfs, if the discrepancy usually observed between these two oxygen abundance indicators in young open clusters is seen to decrease with age in older open clusters.

Finally, we estimate the Fe-core mass fraction of a typical rocky planet in \kep\ system by using the photospheric abundances as proxies for the initial chemical composition of the disk. Our estimate of $\approx$24\% suggests that rocky planets in metal-poor, $\alpha$-enhanced systems may tend to be less dense than rocky planets of similar size in more metal-rich systems.

\acknowledgements{We thank all engineers and technicians involved in PEPSI, in particular our late Emil Popow who passed away much too early, as well as John Little and his LBTO mountain crew. Our thanks also go to Christian Veillet, LBT director, for having us on the radar screen. The many comments by our referee, V.~Adibekyan, are also much appreciated. It led to an improved version of the paper. It is also our honor to thank the German Federal Ministry (BMBF) for the year-long support through their ``ground-based Verbundforschung'' for LBT/PEPSI. }

\appendix

\section{Spectral lines measured}

	\begin{table*}
	\caption{Spectral lines measured, equivalent widths (in m\AA ), and abundances.} \label{tab:lines_kep444}
	\begin{tabular}{lcccllll}
	\hline \noalign{\smallskip}
	 & $\lambda$ & $\chi$ & $\log{gf}$ & The Sun && \kep &  \\
         Ion & (\AA) & (eV) & (-) & EW$_{\odot}$ & $\log{N_{\odot}}$ & EW & $\log{N}$ \\
	\noalign{\smallskip}\hline \noalign{\smallskip}
	\ion{C}{i}  & 5052.17 & 7.69 & -1.304 & 37.2 & 8.53 &  10.0 & 8.22 \\
	\ion{C}{i}  & 5380.34 & 7.69 & -1.615 & 21.9 & 8.52 &  08.0 & 8.45 \\
	\ion{C}{i}  & 7113.18 & 8.65 & -0.685 & 22.9 & 8.49 &  12.1 & 8.71 \\
    $[$\ion{O}{i}$]$ & 6300.30 & 0.00 & -9.717 & 05.3 & 8.66 & 05.4 & 8.54 \\
	\ion{O}{i}  & 7771.94 & 9.15 & +0.369 & 69.3 & 8.72 &  25.2 & 8.59 \\
	\ion{O}{i}  & 7774.17 & 9.15 & +0.223 & 61.4 & 8.74 &  23.1 & 8.68 \\
	\ion{O}{i}  & 7775.39 & 9.15 & +0.001 & 48.7 & 8.77 &  16.9 & 8.70 \\
	\ion{Na}{i} & 5682.63 & 2.10 & -0.700 & 98.0 & 6.27 &  94.6 & 5.74 \\
	\ion{Na}{i} & 6154.23 & 2.10 & -1.560 & 36.6 & 6.27 &  35.2 & 5.89 \\
	\ion{Na}{i} & 6160.75 & 2.10 & -1.260 & 57.0 & 6.29 &  55.4 & 5.87 \\
	\ion{Mg}{i} & 4730.03 & 4.35 & -2.523 & 72.8 & 7.89 &  76.7 & 7.52 \\
	\ion{Mg}{i} & 5711.09 & 4.35 & -1.833 & 105.3 & 7.61 & 126.5 & 7.32 \\
	\ion{Mg}{i} & 6318.72 & 5.11 & -1.945 & 45.5 & 7.57 &  44.6 & 7.29 \\
	\ion{Mg}{i} & 6319.24 & 5.11 & -2.165 & 30.8 & 7.55 &  30.1 & 7.30 \\
	\ion{Mg}{i} & 6965.41 & 5.75 & -1.510 & 24.3 & 7.29 &  24.3 & 7.14 \\
	\ion{Al}{i} & 6696.02 & 3.14 & -1.347 & 37.3 & 6.25 &  46.3 & 6.05 \\
	\ion{Al}{i} & 6698.67 & 3.14 & -1.647 & 21.0 & 6.22 &  27.2 & 6.04 \\
	\ion{Si}{i} & 5690.43 & 4.93 & -1.769 & 53.4 & 7.53 &  34.7 & 7.13 \\
	\ion{Si}{i} & 5701.10 & 4.93 & -1.581 & 38.5 & 7.10 &  26.7 & 6.79 \\
	\ion{Si}{i} & 5772.15 & 5.08 & -1.358 & 52.3 & 7.23 &  33.9 & 6.86 \\
	\ion{Si}{i} & 6125.02 & 5.61 & -1.464 & 31.3 & 7.46 &  17.1 & 7.12 \\
	\ion{Si}{i} & 6142.48 & 5.62 & -1.295 & 34.1 & 7.34 &  19.5 & 7.02 \\
	\ion{Si}{i} & 6145.02 & 5.62 & -1.310 & 38.1 & 7.42 &  21.3 & 7.07 \\
	\ion{Si}{i} & 6243.82 & 5.62 & -1.242 & 46.4 & 7.49 &  27.2 & 7.14 \\
	\ion{Si}{i} & 6244.47 & 5.62 & -1.093 & 45.2 & 7.32 &  26.0 & 6.96 \\
	\ion{Si}{i} & 6414.98 & 5.87 & -1.035 & 47.2 & 7.49 &  26.0 & 7.14 \\
	\ion{Si}{i} & 6741.63 & 5.98 & -1.428 & 16.0 & 7.37 &  04.1 & 6.79 \\
	\ion{Si}{i} & 6848.58 & 5.86 & -1.524 & 17.0 & 7.39 &  07.2 & 7.01 \\
	\ion{Si}{i} & 7405.77 & 5.61 & -0.313 & 90.2 & 7.11 &  65.5 & 6.74 \\
    \ion{K}{i}  & 7698.98 & 0.00 & -0.170 & 155.3 & 5.27 & 185.3 & 4.73 \\
	\ion{Ca}{i} & 5867.56 & 2.93 & -1.570 & 25.5 & 6.33 &  29.6 & 6.02 \\
	\ion{Ca}{i} & 6161.30 & 2.52 & -1.266 & 63.3 & 6.30 &  77.7 & 5.97 \\
	\ion{Ca}{i} & 6166.44 & 2.52 & -1.142 & 69.0 & 6.26 &  83.7 & 5.91 \\
	\ion{Ca}{i} & 6169.04 & 2.52 & -0.797 & 90.9 & 6.26 & 108.1 & 5.85 \\
	\ion{Ca}{i} & 6169.56 & 2.53 & -0.478 & 110.0 & 6.21 & 132.5 & 5.78 \\
	\ion{Ca}{i} & 6455.60 & 2.52 & -1.340 & 55.7 & 6.24 &  67.6 & 5.90 \\
	\ion{Ca}{i} & 6493.78 & 2.52 & -0.109 & 123.8 & 5.98 & 148.6 & 5.53 \\
	\ion{Ca}{i} & 6499.65 & 2.52 & -0.818 & 85.4 & 6.18 &  96.4 & 5.73 \\
	\ion{Sc}{ii}& 6245.64 & 1.51 & -1.030 & 34.3 & 3.06 &  23.9 & 2.73 \\
	\ion{Sc}{ii}& 6604.60 & 1.36 & -1.309 & 35.1 & 3.18 &  24.1 & 2.83 \\
	\ion{Ti}{i} & 5022.87 & 0.83 & -0.434 & 72.5 & 4.79 & 102.4 & 4.44 \\
	\ion{Ti}{i} & 5024.84 & 0.82 & -0.602 & 68.9 & 4.88 &  94.4 & 4.48 \\
	\ion{Ti}{i} & 5866.45 & 1.07 & -0.840 & 46.8 & 4.89 &  76.8 & 4.62 \\
	\ion{Ti}{i} & 6091.17 & 2.27 & -0.423 & 14.5 & 4.93 &  30.4 & 4.78 \\
	\ion{Ti}{i} & 6098.66 & 3.06 & -0.010 & 05.4 & 4.81 &  10.5 & 4.68 \\
	\ion{Ti}{i} & 6258.10 & 1.44 & -0.355 & 51.6 & 4.84 &  77.3 & 4.53 \\
	\ion{Ti}{i} & 6261.10 & 1.43 & -0.479 & 48.8 & 4.90 &  74.5 & 4.60 \\
	\ion{Ti}{ii}& 5154.07 & 1.57 & -1.750 & 72.2 & 5.01 &  77.7 & 4.90 \\
	\ion{Ti}{ii}& 5336.79 & 1.58 & -1.590 & 70.6 & 4.82 &  59.9 & 4.45 \\
	\ion{Ti}{ii}& 5381.02 & 1.57 & -1.920 & 57.3 & 4.87 &  46.8 & 4.55 \\
	\ion{V}{i}  & 5737.06 & 1.06 & -0.740 & 12.0 & 3.98 &  28.5 & 3.75 \\
	\ion{V}{i}  & 6081.44 & 1.05 & -0.579 & 13.8 & 3.85 &  33.9 & 3.66 \\
	\ion{V}{i}  & 6090.21 & 1.08 & -0.062 & 32.7 & 3.85 &  58.3 & 3.57 \\
	\ion{V}{i}  & 6111.65 & 1.04 & -0.715 & 11.5 & 3.89 &  31.5 & 3.74 \\
	\ion{V}{i}  & 6224.53 & 0.29 & -2.010 & 05.6 & 4.08 &  21.7 & 3.97 \\
	\ion{V}{i}  & 6243.11 & 0.30 & -0.980 & 29.3 & 3.90 &  71.0 & 3.78 \\
	\ion{V}{i}  & 6251.83 & 0.29 & -1.340 & 15.4 & 3.89 &  45.9 & 3.75 \\
	\ion{Cr}{i} & 5702.31 & 3.45 & -0.667 & 25.4 & 5.83 &  23.2 & 5.37 \\
	\ion{Cr}{i} & 5783.06 & 3.32 & -0.500 & 32.0 & 5.68 &  30.8 & 5.22 \\
	\ion{Cr}{i} & 5783.85 & 3.32 & -0.295 & 42.3 & 5.67 &  30.8 & 5.20 \\
	\ion{Cr}{i} & 5787.92 & 3.32 & -0.083 & 45.4 & 5.51 &  44.4 & 5.03 \\
	\ion{Cr}{i} & 6330.09 & 0.94 & -2.920 & 26.3 & 5.61 &  40.3 & 5.18 \\
	\ion{Cr}{i} & 7400.25 & 2.90 & -0.111 & 74.2 & 5.54 &  86.1 & 5.11 \\
	\noalign{\smallskip}\hline
	\end{tabular}
	\end{table*}

\setcounter{table}{0}
	\begin{table*}
	\caption{(continued)} \label{tab:lines_kep444_part2}
	\begin{tabular}{lcccllll}
	\hline \noalign{\smallskip}
	 & $\lambda$ & $\chi$ &$\log{gf}$& The Sun && \kep &  \\
         Ion & (\AA) & (eV) & (-) & EW$_{\odot}$ & $\log{N_{\odot}}$ & EW & $\log{N}$ \\
	\noalign{\smallskip}\hline \noalign{\smallskip}
	\ion{Fe}{i} &  4779.44 &  3.42 &  -2.000 &  40.5 &  7.25 &  39.0 &  6.77 \\
        \ion{Fe}{i} &  4788.76 &  3.24 &  -1.800 &  65.5 &  7.29 &  62.2 &  6.69 \\
        \ion{Fe}{i} &  5054.64 &  3.64 &  -1.900 &  39.8 &  7.33 &  38.3 &  6.88 \\
        \ion{Fe}{i} &  5322.04 &  2.28 &  -2.800 &  59.8 &  7.24 &  58.8 &  6.61 \\
        \ion{Fe}{i} &  5379.57 &  3.69 &  -1.500 &  60.6 &  7.33 &  53.6 &  6.75 \\
        \ion{Fe}{i} &  5522.45 &  4.21 &  -1.600 &  42.3 &  7.52 &  33.5 &  7.00 \\
        \ion{Fe}{i} &  5543.94 &  4.22 &  -1.100 &  61.5 &  7.46 &  52.7 &  6.91 \\
        \ion{Fe}{i} &  5546.50 &  4.37 &  -1.300 &  50.6 &  7.58 &  41.3 &  7.06 \\
        \ion{Fe}{i} &  5546.99 &  4.22 &  -1.900 &  27.1 &  7.58 &  24.8 &  7.20 \\
        \ion{Fe}{i} &  5560.21 &  4.43 &  -1.200 &  51.3 &  7.53 &  39.7 &  6.98 \\
        \ion{Fe}{i} &  5587.57 &  4.14 &  -1.900 &  38.9 &  7.69 &  32.1 &  7.20 \\
        \ion{Fe}{i} &  5646.68 &  4.26 &  -2.500 &   7.4 &  7.53 &  05.0 &  7.04 \\
        \ion{Fe}{i} &  5651.47 &  4.47 &  -2.000 &  18.9 &  7.70 &  12.7 &  7.20 \\
        \ion{Fe}{i} &  5652.32 &  4.26 &  -1.900 &  26.1 &  7.64 &  18.3 &  7.11 \\
        \ion{Fe}{i} &  5661.35 &  4.28 &  -1.700 &  22.3 &  7.35 &  15.4 &  6.84 \\
        \ion{Fe}{i} &  5667.52 &  4.18 &  -1.600 &  50.4 &  7.66 &  36.2 &  7.04 \\
        \ion{Fe}{i} &  5679.02 &  4.65 &  -0.900 &  58.6 &  7.58 &  46.9 &  7.04 \\
        \ion{Fe}{i} &  5680.24 &  4.19 &  -2.600 &  11.0 &  7.73 &  07.8 &  7.25 \\
        \ion{Fe}{i} &  5731.76 &  4.26 &  -1.300 &  56.9 &  7.57 &  46.0 &  7.01 \\
        \ion{Fe}{i} &  5741.85 &  4.26 &  -1.900 &  31.2 &  7.64 &  23.4 &  7.14 \\
        \ion{Fe}{i} &  5752.03 &  4.55 &  -1.200 &  53.8 &  7.66 &  42.7 &  7.13 \\
        \ion{Fe}{i} &  5775.08 &  4.22 &  -1.300 &  57.9 &  7.55 &  47.3 &  6.98 \\
        \ion{Fe}{i} &  5778.45 &  2.59 &  -3.500 &  22.0 &  7.44 &  21.2 &  6.92 \\
        \ion{Fe}{i} &  5809.22 &  3.88 &  -1.800 &  48.3 &  7.59 &  39.1 &  7.03 \\
        \ion{Fe}{i} &  6079.00 &  4.65 &  -1.100 &  45.9 &  7.55 &  32.4 &  7.00 \\
        \ion{Fe}{i} &  6085.26 &  2.76 &  -3.100 &  42.5 &  7.64 &  59.1 &  7.39 \\
        \ion{Fe}{i} &  6098.24 &  4.56 &  -1.900 &  16.5 &  7.57 &  10.0 &  7.05 \\
        \ion{Fe}{i} &  6151.62 &  2.18 &  -3.300 &  48.3 &  7.37 &  49.6 &  6.80 \\
        \ion{Fe}{i} &  6159.37 &  4.61 &  -2.000 &  12.2 &  7.56 &  07.3 &  7.04 \\
        \ion{Fe}{i} &  6165.36 &  4.14 &  -1.500 &  43.7 &  7.37 &  33.8 &  6.83 \\
        \ion{Fe}{i} &  6187.99 &  3.94 &  -1.700 &  46.1 &  7.47 &  36.6 &  6.92 \\
        \ion{Fe}{i} &  6220.78 &  3.88 &  -2.500 &  19.1 &  7.58 &  14.0 &  7.06 \\
        \ion{Fe}{i} &  6226.73 &  3.88 &  -2.200 &  28.2 &  7.56 &  20.5 &  7.02 \\
        \ion{Fe}{i} &  6229.23 &  2.85 &  -2.800 &  35.7 &  7.30 &  32.7 &  6.76 \\
        \ion{Fe}{i} &  6240.65 &  2.22 &  -3.200 &  47.3 &  7.31 &  47.6 &  6.74 \\
        \ion{Fe}{i} &  6293.92 &  4.84 &  -1.700 &  13.2 &  7.56 &  07.4 &  7.04 \\
        \ion{Fe}{i} &  6380.74 &  4.19 &  -1.400 &  51.2 &  7.46 &  40.3 &  6.90 \\
        \ion{Fe}{i} &  6392.54 &  2.28 &  -4.000 &  17.7 &  7.53 &  19.3 &  7.04 \\
        \ion{Fe}{i} &  6597.56 &  4.79 &  -1.100 &  43.9 &  7.57 &  31.6 &  7.06 \\
        \ion{Fe}{i} &  6608.02 &  2.28 &  -4.000 &  17.0 &  7.49 &  19.6 &  7.04 \\
        \ion{Fe}{i} &  6627.54 &  4.55 &  -1.700 &  27.1 &  7.62 &  19.3 &  7.14 \\
        \ion{Fe}{i} &  6703.57 &  2.76 &  -3.200 &  35.9 &  7.54 &  34.3 &  7.02 \\
        \ion{Fe}{i} &  6710.32 &  1.49 &  -4.900 &  15.1 &  7.49 &  18.1 &  6.96 \\
        \ion{Fe}{i} &  6713.74 &  4.80 &  -1.600 &  20.7 &  7.62 &  13.0 &  7.13 \\
        \ion{Fe}{i} &  6716.22 &  4.58 &  -1.900 &  15.5 &  7.58 &  06.2 &  6.87 \\
        \ion{Fe}{i} &  6725.35 &  4.10 &  -2.300 &  17.9 &  7.57 &  11.5 &  7.02 \\
        \ion{Fe}{i} &  6726.67 &  4.61 &  -1.100 &  25.8 &  7.10 &  32.9 &  6.95 \\
        \ion{Fe}{i} &  6733.15 &  4.64 &  -1.600 &  25.9 &  7.58 &  16.2 &  7.04 \\
        \ion{Fe}{i} &  6739.52 &  1.56 &  -4.800 &  11.3 &  7.32 &  14.4 &  6.83 \\
        \ion{Fe}{i} &  6752.72 &  4.64 &  -1.300 &  35.1 &  7.49 &  23.6 &  6.96 \\
        \ion{Fe}{ii} &  4620.52 &  2.83 &  -3.300 &  51.3 &  7.39 &  23.9 &  6.87 \\
        \ion{Fe}{ii} &  5197.58 &  3.23 &  -2.300 &  79.0 &  7.36 &  44.5 &  6.77 \\
        \ion{Fe}{ii} &  5234.62 &  3.22 &  -2.300 &  83.3 &  7.37 &  47.0 &  6.74 \\
        \ion{Fe}{ii} &  5264.81 &  3.23 &  -3.100 &  47.1 &  7.47 &  20.3 &  7.00 \\
        \ion{Fe}{ii} &  5414.07 &  3.22 &  -3.600 &  28.7 &  7.56 &  09.4 &  7.09 \\
        \ion{Fe}{ii} &  5425.26 &  3.20 &  -3.400 &  41.0 &  7.56 &  19.3 &  7.19 \\
        \ion{Fe}{ii} &  6149.26 &  3.89 &  -2.800 &  36.2 &  7.55 &  10.6 &  7.02 \\
        \ion{Fe}{ii} &  6247.56 &  3.89 &  -2.400 &  51.7 &  7.47 &  19.0 &  6.94 \\
        \ion{Fe}{ii} &  7711.72 &  3.90 &  -2.700 &  45.3 &  7.54 &  13.8 &  7.00 \\
	\ion{Co}{i} & 5352.04 & 3.58 & +0.060 & 25.1 & 4.57 & 22.5 & 4.19\\
	\ion{Co}{i} & 5301.04 & 1.71 & -2.000 & 20.8 & 4.97 &  30.9 & 4.66 \\
	\ion{Co}{i} & 5647.23 & 2.28 & -1.560 & 14.4 & 4.88 &  19.5 & 4.56 \\
	\ion{Co}{i} & 6093.14 & 1.74 & -2.440 & 09.6 & 4.99 &  14.6 & 4.68 \\
	\ion{Co}{i} & 6814.94 & 1.96 & -1.900 & 19.3 & 4.98 &  27.4 & 4.67 \\
	\noalign{\smallskip}\hline \noalign{\smallskip}
	\end{tabular}
	\end{table*}

\setcounter{table}{0}
	\begin{table*}
	\caption{(continued)} \label{tab:lines_kep444_part3}
	\begin{tabular}{lcccllll}
	\hline \noalign{\smallskip}
	 & $\lambda$ & $\chi$ &$\log{gf}$& The Sun && \kep &  \\
         Ion & (\AA) & (eV) & (-) & EW$_{\odot}$ & $\log{N_{\odot}}$ & EW & $\log{N}$ \\
	\noalign{\smallskip}\hline \noalign{\smallskip}
	\ion{Ni}{i} & 5748.35 & 1.68 & -3.260 & 28.5 & 6.19 &  28.6 & 5.71 \\
	\ion{Ni}{i} & 5754.66 & 1.94 & -2.330 & 76.6 & 6.44 &  74.5 & 5.83 \\
	\ion{Ni}{i} & 5760.83 & 4.11 & -0.800 & 35.1 & 6.24 &  23.4 & 5.75 \\
	\ion{Ni}{i} & 5846.99 & 1.68 & -3.210 & 22.7 & 6.00 &  22.2 & 5.51 \\
	\ion{Ni}{i} & 6108.11 & 1.68 & -2.450 & 64.6 & 6.04 &  63.6 & 5.47 \\
	\ion{Ni}{i} & 6111.07 & 4.09 & -0.870 & 34.2 & 6.26 &  22.5 & 5.77 \\
	\ion{Ni}{i} & 6128.96 & 1.68 & -3.330 & 24.8 & 6.15 &  24.4 & 5.67 \\
	\ion{Ni}{i} & 6130.13 & 4.27 & -0.960 & 21.8 & 6.24 &  12.6 & 5.74 \\
	\ion{Ni}{i} & 6133.96 & 4.09 & -1.830 & 05.8 & 6.27 &  04.3 & 5.92 \\
	\ion{Ni}{i} & 6175.36 & 4.09 & -0.559 & 48.2 & 6.21 &  34.4 & 5.70 \\
	\ion{Ni}{i} & 6176.81 & 4.09 & -0.260 & 62.9 & 6.16 &  47.3 & 5.61 \\
	\ion{Ni}{i} & 6177.24 & 1.83 & -3.500 & 14.8 & 6.19 &  13.6 & 5.70 \\
	\ion{Ni}{i} & 6186.71 & 4.11 & -0.960 & 30.8 & 6.29 &  19.1 & 5.78 \\
	\ion{Ni}{i} & 6204.60 & 4.09 & -1.100 & 21.2 & 6.19 &  13.1 & 5.71 \\
	\ion{Ni}{i} & 6223.98 & 4.11 & -0.910 & 27.5 & 6.17 &  17.3 & 5.68 \\
	\ion{Ni}{i} & 6230.09 & 4.11 & -1.260 & 21.0 & 6.36 &  13.2 & 5.90 \\
	\ion{Ni}{i} & 6327.59 & 1.68 & -3.150 & 38.2 & 6.25 &  38.6 & 5.76 \\
	\ion{Ni}{i} & 6370.34 & 3.54 & -1.940 & 12.8 & 6.23 &  09.1 & 5.80 \\
	\ion{Ni}{i} & 6378.25 & 4.15 & -0.830 & 31.9 & 6.22 &  20.7 & 5.74 \\
	\ion{Ni}{i} & 6635.12 & 4.42 & -0.820 & 25.4 & 6.31 &  15.8 & 5.86 \\
	\ion{Ni}{i} & 6643.63 & 1.68 & -2.300 & 93.9 & 6.38 &  93.9 & 5.74 \\
	\ion{Ni}{i} & 6767.77 & 1.83 & -2.170 & 78.4 & 6.11 &  76.2 & 5.50 \\
	\ion{Ni}{i} & 6842.04 & 3.66 & -1.480 & 26.4 & 6.26 &  18.3 & 5.79 \\
    \ion{Mn}{i} & 5432.55 & 0.00 & -4.382 & 52.2 & 5.34 &  79.6 & 4.83 \\
    \ion{Mn}{i} & 6013.51 & 3.07 & -0.765 & 87.2 & 5.36 &  81.3 & 4.76 \\
    \ion{Mn}{i} & 6021.82 & 3.08 & -0.537 & 95.1 & 5.63 &  88.5 & 4.90 \\
    \ion{Zn}{i} & 4722.16 & 4.03 & -0.338 & 67.9 & 4.43 &  52.2 & 4.02 \\
    \ion{Zn}{i} & 4810.53 & 4.08 & -0.137 & 71.9 & 4.35 &  56.9 & 3.94 \\
    \ion{Y}{ii} & 5200.40 & 0.99 & -0.570 & 38.0 & 2.11 &  22.6 & 1.57 \\
	\noalign{\smallskip}\hline \noalign{\smallskip}
	\end{tabular}
	\end{table*}

\end{document}